%% file: aistat_preprint.tex
\documentclass[twoside]{article}
\pdfoutput=1
\usepackage[accepted]{aistats2023}
\usepackage{amsmath}
\usepackage[pdftex]{graphicx}
\usepackage{psfrag,epsf}
\usepackage{enumerate}
\usepackage{xfrac}
\usepackage{breqn}
\usepackage{tcolorbox}
\usepackage[export]{adjustbox}

\allowdisplaybreaks

\usepackage{enumerate}
\usepackage{url} % not crucial - just used below for the URL 
\usepackage{float}
 \usepackage[T1]{fontenc}
\RequirePackage{fix-cm}
\let\accentvec\vec
\makeatletter
\let\@fnsymbol\@arabic
\makeatother

\usepackage[final]{pdfpages}
\usepackage{subfig}

\usepackage{epsfig}
\usepackage{amssymb}% http://ctan.org/pkg/amssymb
\usepackage{bm}
\usepackage{pifont}% http://ctan.org/pkg/pifont
\usepackage[misc]{ifsym}
\usepackage{setspace}
\usepackage{epstopdf}
\usepackage{secdot}
\usepackage{cancel}

\let\vec\accentvec
\usepackage{mathtools}
\usepackage{dsfont}
\usepackage{siunitx}
\usepackage{latexsym}
\usepackage{mathrsfs}
\usepackage{color}
\usepackage{xcolor}
\usepackage{soul}
\usepackage{setspace}
\usepackage{amsthm}
\usepackage{parskip}
\usepackage[ruled]{algorithm2e}
\usepackage[colorlinks = true, pdfstartview = FitV, linkcolor = blue, citecolor = blue, urlcolor = blue, bookmarks = false]{hyperref}
\usepackage[authoryear, round]{natbib}
\usepackage{todonotes}

\newcommand{\m}[1]{\mathrm{#1} }
\renewcommand{\cal}[1]{\mathcal{#1}}
\renewcommand{\v}[1]{\boldsymbol{#1}}
\newcommand{\bb}[1]{\mathbb{#1}}

\usepackage{xcolor, color}
\definecolor{red}{rgb}{0.81, 0.09, 0.13}

\newtheorem{prop1}{Proposition}
\begin{document}
\include{aistat_body_preprint}
\include{supplement_body}

\end{document}

%% file: aistat_body_preprint.tex
\newtheorem{corollary}{Corollary}
\renewcommand\refname{}

\twocolumn[

\aistatstitle{Transport Reversible Jump Proposals}

\aistatsauthor{Laurence Davies\textsuperscript{1,3} \And Robert Salomone\textsuperscript{2,3} \And Matthew Sutton\textsuperscript{1,3} \And Christopher Drovandi\textsuperscript{1,3}}

 \runningauthor{Laurence Davies, Robert Salomone, Matthew Sutton, Christopher Drovandi}

\aistatsaddress{\textsuperscript{1} School of Mathematical Sciences, Queensland University of Technology \\ \textsuperscript{2} School of Computer Science, Queensland University of Technology \\  \textsuperscript{3} Centre for Data Science, Queensland University of Technology} 
]

\begin{abstract}
Reversible jump Markov chain Monte Carlo (RJMCMC) proposals that achieve reasonable acceptance rates and mixing are notoriously difficult to design in most applications. Inspired by recent advances in deep neural network-based normalizing flows and density estimation, we demonstrate an approach to enhance the efficiency of RJMCMC sampling by performing transdimensional jumps involving reference distributions. In contrast to other RJMCMC proposals, the proposed method is the first to apply a non-linear transport-based approach to construct efficient proposals between models with complicated dependency structures. It is shown that, in the setting where exact transports are used, our RJMCMC proposals have the desirable property that the acceptance probability depends only on the model probabilities. Numerical experiments demonstrate the efficacy of the approach.
\end{abstract}

\section{INTRODUCTION}
\label{sec:intro}
The problem of interest is sampling from a probability distribution $\pi$ with support  
\begin{align}\mathcal{X}&=\bigcup_{k\in\mathcal{K}}(\{k\}\times\Theta_k)\label{eq:jointspace},\end{align}
where $\mathcal{K}$ is a discrete {\em index} set, and $\Theta_k\subseteq\mathbb{R}^{n_k}$, where the $n_k$ may differ, and hence $\mathcal{X}$ is a {\em transdimensional} space. The choice of notation $\Theta_k$ is made as the problem typically arises in Bayesian model selection, where such sets correspond to the space of model {\em parameters}, and $k\in\mathcal{K}$ is a {\em model index} or indicator.
The reversible jump Markov chain Monte Carlo (RJMCMC) algorithm, formally introduced by \cite{green_reversible_1995}, generalizes the Metropolis--Hastings algorithm \cite[]{hastings_monte_1970}  via the introduction of user-specified diffeomorphisms\footnote{bijective functions that are differentiable and have a differentiable inverse} $h_{k,k'}:\Theta_k\times\mathcal{U}_k\rightarrow\Theta_{k'}\times\mathcal{U}_{k'}$ where $\mathcal{U}_k, \mathcal{U}_{k'}$ are (possibly empty) sets chosen to ensure dimensions are matched. Due to this additional complexity, RJMCMC proposals that perform well in practice are generally challenging to design. 

RJMCMC methods are a frequently revisited research topic for which many approaches exist. \cite{brooks_efficient_2003} outline several approaches to improve the efficiency of jump proposals on various problem types, including variable selection and nested models. \cite{green_delayed_2001} identify the efficiency issues associated with na\"ive transdimensional proposals between non-overlapping targets and propose a delayed-rejection auxiliary-proposal mechanism. \cite{al-awadhi_improving_2004} employ an auxiliary target density instead of an auxiliary proposal. \cite{hastie_towards_2005} proposes an approach for (potentially) multi-modal conditional target densities, achieved by
fitting a Gaussian mixture model to each individual conditional target, and using a shift and whitening transformation corresponding to a randomly selected mixture component.
\cite{fan_automating_2008} propose a conditional factorization of a differentiable target density to sequentially construct a proposal density. \cite{karagiannis_annealed_2013} propose the construction of a Markov chain through an annealed sequence of intermediate distributions between models to encourage higher acceptance rates between model jumps. \cite{farr_efficient_2015} propose a KD-tree approximation of the target density for auxiliary variable draws to improve the efficiency of RJMCMC proposals. \cite{gagnon_informed_2021} uses the locally-informed MCMC approach for discrete spaces developed by \cite{zanella_informed_2017} to improve the exploration efficiency of the model space when global jump proposals are unavailable. However, there appears to be no {\em general} strategy for the design of across-model proposals that is widely applicable, theoretically justified, and of practical use.

The use of measure transport to enhance sampling methods is an area of recent interest. This was first formalized and demonstrated in \cite{parno_transport_2018}, where \emph{approximate} transport maps (TMs) are used for accelerating MCMC. The form of the approximate transports as applied to MCMC sampling is described in general terms, but their choice of transports uses systems of orthogonal multivariate polynomials. 
The application of approximate TMs to enhance sampling methods includes mapping a deterministic step within sequential Monte Carlo \cite[]{arbel_annealed_2021}; the transformation of a continuous target distribution to an easier-to-sample distribution via a map learned using stochastic variational inference \cite[]{hoffman_neutra-lizing_2019};
and the use of approximate TMs for the construction of independence proposals in an adaptive MCMC algorithm \cite[]{gabrie_adaptive_2022}. 

However, despite the considerable promise of incorporating approximate TMs into sampling methodology, to our knowledge, such ideas have not been considered in the {\em transdimensional} sampling setting.  Such an omission is somewhat surprising, as RJMCMC samplers are constructed using a form of invertible maps involving distributions, and hence it would intuitively appear natural that distributional transport could feature in a useful capacity. This work aims to develop such an approach and to demonstrate its benefits.

\paragraph{Contribution.}
The primary contributions of this work are as follows:

\textbf{I.} A new class of RJMCMC proposals for across-model moves, called {\em transport reversible jump} (TRJ) proposals are developed. In the idealized case where exact transports are used, the proposals are shown to have a desirable property (Proposition \ref{prop1}). \\
\textbf{II.} A numerical study is conducted on challenging examples demonstrating the efficacy of the proposed approach in the setting where {\em approximate} transport maps are used. \\
\textbf{III.} An alternative ``all-in-one'' approach to training approximate TMs is developed, which involves combining  a saturated state space formulation of the target distribution with {\em conditional} TMs.

Code for the numerical experiments is made available at \url{https://github.com/daviesl/trjp}.
\paragraph{Structure of this Article.}
The remainder of this article is structured as follows: Section \ref{sec:background} discusses the required background concepts regarding RJMCMC,  transport maps, and flow-based models. Section \ref{sec:rjrd} introduces a general strategy for using transport maps within RJMCMC and discusses its properties. Section \ref{sec:numerical} conducts a numerical study to demonstrate the efficacy of the strategy in the case where {\em approximate} transports are used. Section \ref{sec:cnfrj} explores an alternative ``all-in-one'' approach to training transport maps and provides an associated numerical example. Section \ref{sec:discussion} concludes the paper.

\paragraph{Notation.} For a function $T$ and distribution $\nu$,  $T{\sharp}\nu$ denotes the {\em pushforward} of $\nu$ under $T$. That is, if $\v Z \sim \nu$, then $T\sharp \nu$ is the probability distribution of $T(\v Z)$. For a univariate probability distribution $\nu$, define 
$\otimes_n \nu$ as $\underbrace{\nu \otimes \cdots  \otimes \nu}_{n \text{ times}}.$ 
Throughout, univariate functions are to be interpreted as applied element-wise when provided with a vector as input. The symbol $\odot$ denotes the Hadamard (element-wise) product. The univariate standard normal probability density function is denoted as $\phi$, $\phi_{d}$ is the $d$-dimensional multivariate standard normal probability density function, and $\phi_{\m\Sigma_{d\times d}}$ is the $d$-dimensional multivariate normal probability density function centered at $\v 0_d$ with covariance %$\boldsymbol{\Sigma}_{d}$
$\m \Sigma_{d\times d}$. For a function $f:\mathbb{R}^n\rightarrow\mathbb{R}^n$, the notation $|J_f(\v \theta)|$ denotes the absolute value of the determinant of the Jacobian matrix of $f$ evaluated at some $\v \theta \in \bb R^n$. For distributions $\pi$ defined on sets of the form in \eqref{eq:jointspace}, we write $\pi_k$ for the distribution conditional on $k$, and $\pi(k)$ for its $k$-marginal distribution. 

\section{BACKGROUND}\label{sec:background}
\subsection{Reversible Jump Markov Chain Monte Carlo}\label{sec:rjmcmc}
For a distribution $\pi$ defined on a space of the form in \eqref{eq:jointspace}, with associated probability density function $\pi(\v x)$, the standard method to construct a $\pi$-invariant Metropolis--Hastings kernel (and thus an associated MCMC sampler) is the {\em reversible jump} approach introduced in the seminal work by \cite{green_reversible_1995}.  
The proposal mechanism is constructed to take $\v x=(k,\v \theta_k)$ to $\v x'=(k', \v \theta_{k'}')$ where the dimensions of $\v \theta_k$ and $\v \theta_{k'}'$ are $n_k$ and $n_{k'}$, respectively. The approach employs {\em dimension matching}, introducing auxiliary random variables $\v u_k\sim g_{k,k'}$ and $\v u_{k'}'\sim g_{k',k}$ of dimensions $w_k$ and $w_{k'}$, which are arbitrary provided that $n_k + w_k = n_{k'} + w_{k'}$. A proposal is then made using these auxiliary random variables and a chosen diffeomorphism $h_{k,k'}$ defined so that $(\v\theta_{k'}', \v u_{k'}')=h_{k,k'}(\v\theta_k,\v u_k)$. A discrete proposal distribution $j_{k}$ is also specified for each $k\in \mathcal{K}$, where $j_k(k')$ defines the probability of proposing to model $k'$ from model $k$. More generally, the distributions $j_k$ may also depend on $\v\theta_k$, but we do not consider this case. With the above formulation of the proposal, the RJMCMC acceptance probability is 
\begin{align}
    \alpha(\v x,\v x')&=1\wedge\frac{\pi(\v x')j_{k'}(k)g_{k',k}(\v u_{k'}')}{\pi(\v x)j_{k}(k')g_{k,k'}(\v u_k)}\big|J_{h_{k,k'}}(\v\theta_k,\v u_k)\big|.\label{eq:rjacceptanceprobability}
\end{align}

\subsection{Transport Maps and Flow-Based Models}\label{sec:flows}
Consider two random vectors $\v \theta\sim\mu_{\v \theta}$ and $\v Z\sim\mu_{\v z}$, such that their distributions $\mu_{\v \theta}$ and $\mu _{\v z}$ are absolutely continuous with respect to $n$-dimensional Lebesgue measure. A function $T:\mathbb{R}^n\rightarrow\mathbb{R}^n$ is called a \emph{transport map} (TM) from $\mu_{\v \theta}$ to $\mu_{\v z}$ if $\mu_{\v z}=T\sharp\mu_{\v \theta}\label{eq:tmdef}$.
In this setting, we refer to $\mu_{\v \theta}$ as the \emph{target} distribution and $\mu_{\v z}$ as the \emph{reference} distribution. Transport maps between two prescribed distributions are known to exist under mild conditions (see e.g., \cite{parno_transport_2018} and the references therein). 

One strategy for obtaining approximate TMs from samples of a target $\pi$ is via density estimation with a family of distributions arising from transformations. Let $\{T_{\v \psi}\}$ be a family of diffeomorphisms parameterized by $\v \psi \in \Psi$ with domain on the support of some arbitrary {\em base} distribution $\mu_{\v z}$. Then, for fixed $\v \psi$, the probability density function of the random vector $\v \zeta = T_{\v \psi}(\v Z)$ is 
\begin{align}
    \mu_{\v \zeta}(\v\zeta ; \v \psi)&=\mu_{\v z}(T_{\v \psi}^{-1}(\v\zeta))|J_{T_{\v \psi}^{-1}}(\v \zeta)|,\,\,\v\zeta\in\bb R^n \label{eq:flowdist}.
\end{align}
An approximate TM from $\mu_{\v z}$ to some prescribed target distribution $\pi$ can be learned as a result of fitting a model of the above form to samples from $\pi$ by minimizing KL divergence between the model and an empirical distribution of possibly-approximate samples from $\pi$, which is equivalent to maximum likelihood estimation.

Families of distributions arising from the specification of highly flexible and scalable  $\{T_{\v \psi}\}$ have been an area of intense recent interest in the machine learning literature, often referred to as {\em flow-based models}, with the families $\{T_{\v \psi}\}$ often referred to as {\em flows} or {\em normalizing flows} (NF). A general class of transform is {\em autoregressive flows}, briefly described as follows. Let $\tau(\cdot; \v \omega)$, called the {\em transformer}, be a univariate diffeomorphism parametrized by $\v \omega \in \v \Omega$, where $\v\Omega$ is the set of admissible parameters. Then, the transformation is defined elementwise via
\begin{equation}
 T(\v Z)_i = \tau(Z_i ; \v \zeta_i(\v z_{< i}; \v \psi)), \quad i=1,\ldots, n, \label{eq:generalARflow} 
\end{equation}
where the functions $\{\v \zeta_i: \bb R^{i-1} \to \v \Omega\}$ are called the {\em conditioners}. In practice, the $\{\v \zeta_i\}$ are each individually subsets of the output of a single neural network that takes $\v Z$ as input and is designed to respect the autoregressive structure. When fitting approximate transports to samples in our experiments in Section \ref{sec:numerical}, for additional flexibility the flow-based model used arises from {\em several} chained transformations of the form in \eqref{eq:generalARflow}, each with different parameters, and where the transformer is a piecewise function involving monotonic rational-quadratic splines \citep{durkan_neural_2019}. For further details and a comprehensive overview of other flows, see \cite{papamakarios_normalizing_2021}.

\section{TRANSPORT REVERSIBLE JUMP PROPOSALS}
\label{sec:rjrd}
\begin{figure}[t]
\vspace{.2in}
\centering
\includegraphics[width=0.45\textwidth,trim={0 0 10 0},clip]{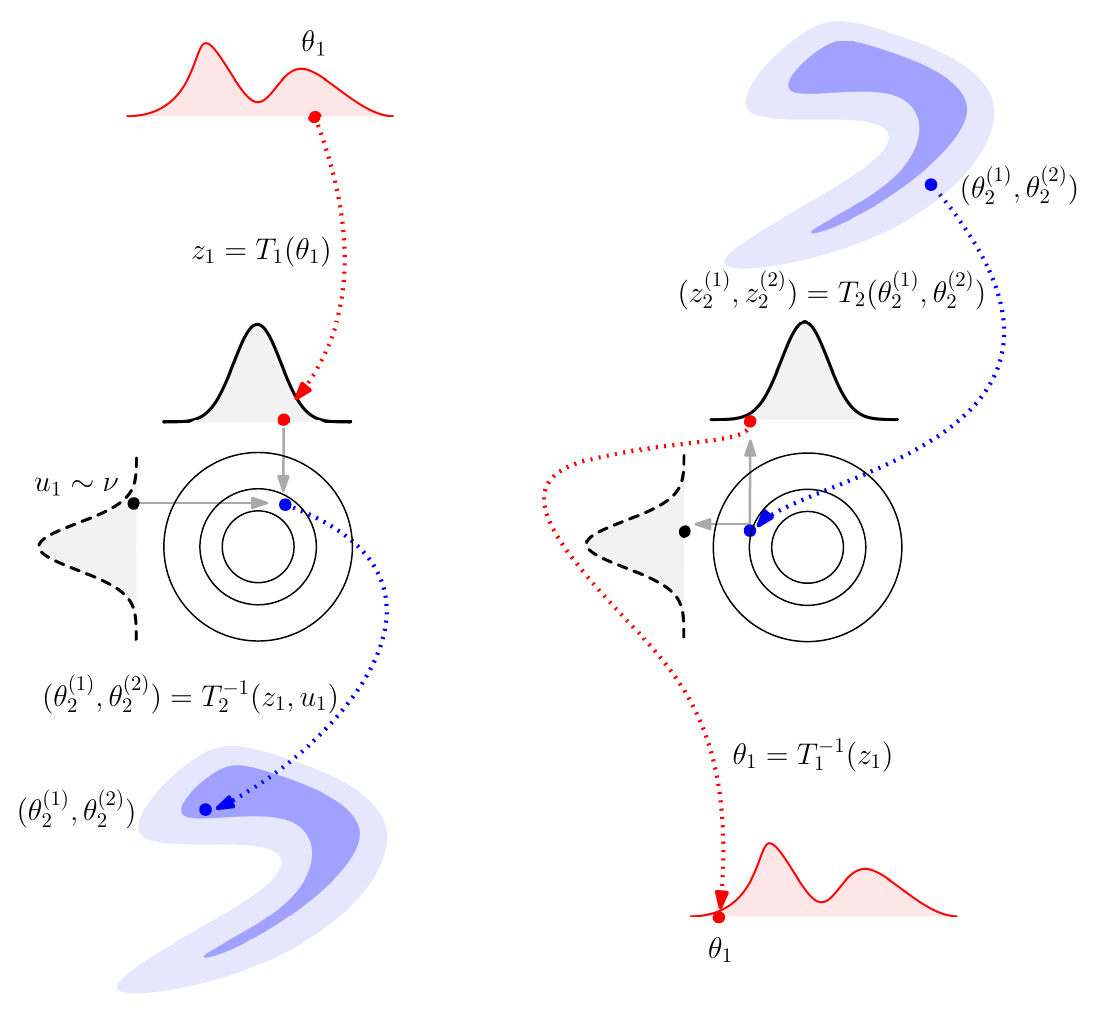}
\caption{Illustration of the proposal class. Here, the reference $\nu$ is Gaussian, and the functions $\bar{h}_{k,k'}$ are the identity map. 
} \label{fig:MainIdea}
\end{figure}

We introduce a special form of general proposal between two models, corresponding to indices $k$ and $k'$. Let $\nu$ be some {\em univariate} reference distribution, and let $\{T_k: \mathbb{R}^{n_k} \to \mathcal{Z}^{n_k} \}$ be a collection of diffeomorphisms such that 
$T_k\sharp \pi_k \approx \otimes_{n_k}\nu, \quad  k\in \mathcal{K}$,
i.e., $T_k$ is an approximate TM from the conditional target corresponding to index $k$, to an appropriately-dimensioned product of independent reference distributions. Similarly, $T_k^{-1}$ is an approximate TM {\em from} the reference to the conditional target for index $k$. 
The general idea of the proposal scheme from model $k$ to model $k'$ is to first apply the transformation that would (approximately) transport $\pi_k$ to its product reference, drawing auxiliary variables (if proposing to a higher-dimensional model), optionally applying a product reference measure preserving diffeomorphism, discarding auxiliary variables (if proposing to a lower dimensional model), and then applying the transformation that would (approximately) transport the augmented vector to the new conditional target. Figure \ref{fig:MainIdea} illustrates this idea for jumps between targets of one and two dimensions.

Formally, we first restrict the dimension of proposed auxiliary variables $\v u$ (if any) to $w_k$, which is defined to be the dimension difference between $\v\theta_k$ and $\v\theta_{k'}'$, i.e., $n_k+w_k=n_{k'}$. Then, assuming $w_k \ge 0$ (a jump is proposed to a model of equal or higher dimension),  the proposal is obtained via
\begin{align}
    \begin{split}
        \v z_k&=T_{k}(\v\theta_k),\\
    \v z_{k'}&=\bar{h}_{k,k'}(\v z_k,\v u_k), \quad \text{where } \v u_k {\sim} \otimes_{w_k}\nu, \\
    \v\theta_{k'}' &= T^{-1}_{k'}(\v z_{k'})
    \label{eq:chaintrans},
    \end{split}
\end{align}
where each 
$\bar{h}_{k,k'}: \mathbb{R}^{n_k} \times \mathbb{R}^{w_k} \rightarrow \mathbb{R}^{n_{k'}}$, 
is a diffeomorphism that both satisfies the pushforward-invariance condition
\begin{align}
    \begin{split}
    \bar{h}_{k, k'}\sharp \otimes_{\max\{ n_k, n_k'\}}\nu &= \otimes_{\max\{ n_k, n_k'\}} \nu,
    \end{split}
\end{align}
and is volume-preserving  (i.e., the absolute value of the Jacobian determinant is always equal to one).  The default choice is to set each $\bar{h}_{k,k'}$ equal to the identity (resulting in the concatenation of the two inputs), but  other choices are possible. For example, as $\otimes_{\max\{ n_k, n_k'\}} \nu$ is exchangeable, permutations are also suitable.
Proposal moves to lower dimensions are similar to those in  \eqref{eq:chaintrans}, but auxiliary variables of appropriate dimension are discarded as opposed to being drawn. The procedure of using our TRJ proposals within RJMCMC is given in Algorithm \ref{algo:TRJ}. 

\begin{algorithm}[ht]
\SetKwInOut{KwInput}{input}
\SetKwInOut{KwOutput}{output}
\small
\DontPrintSemicolon
\caption{TRJ: Transport Reversible Jump MCMC}\label{algo:TRJ}
\KwInput{$\v x= (k,\v\theta_k),
 \{T_k\},
\pi,
\{j_{k}\}, 
\nu, \{P_{k}(\v\theta, d\v\theta')\}$ such that each $P_k$ is $\pi_k$-reversible.}
Draw $k' \sim j_k$\\
\uIf{$k'=k$}{

\text{Draw} $\v\theta_k'\sim P_{k}(\v\theta_k, d\v\theta_k')$ \text{ and } \KwRet{$(k,\v \theta_k')$} 

}\uElse{
$\v z_k\gets T_k(\v\theta_k)$\\
\uIf{$n_{k'}>n_k$}{
$\begin{aligned}
w_k&\gets n_{k'}-n_k\\
\text{Draw } \v u&\sim \otimes_{w_k}\nu\\
g_u &\gets (\otimes_{w_k}\nu)(\v u), \, 
g_u'\gets 1\\
\v z_{k'}&\gets  \bar{h}_{k,k'}(\v z_k, \v u) \\
\end{aligned}$
}
\uElseIf{$n_{k'}<n_k$}{
$\begin{aligned}
w_{k'}&\gets n_{k}-n_{k'}\\
(\v z_{k'}, \v u)&\gets \bar{h}_{k,k'}^{-1}(\v z_k)\\
g_u&\gets 1, \ g_u'\gets (\otimes_{w_{k'}}\nu)(\v u)\\
\end{aligned}$
}
\uElse{
$\begin{aligned}
\v z_{k'}&\gets \bar{h}_{k,k'}^{-1}(\v z_k)\\
g_u&\gets 1, \ g_u'\gets 1\\
\end{aligned}$
}
$\begin{aligned}
    \v \theta_{k'}'&\gets T_{k'}^{-1}(\v z_{k'})\\
    \v x'&\gets(k',\v\theta_{k'}')\\
\alpha&\gets 1 \wedge \frac{\pi(\v x')}{\pi(\v x)}\frac{j_{k'}(k)}{j_{k}(k')}\frac{g_u'}{g_u}
    \Big|J_{T_k}(\v\theta_k)\Big|\ \Big|J_{T_{k'}}(\v \theta_{k'}')\Big|^{-1}
\end{aligned}$\\
Draw $V\sim \mathcal{U}(0,1)$\\ 
\lIf{$\alpha>V$}{
\KwRet{$\v x'$}
}\lElse{
\KwRet{$\v x$}
}
} 
\end{algorithm}

The following proposition provides a formal justification for the proposed approach, showing that when exact transports are used, 
\eqref{eq:rjacceptanceprobability} reduces to a form depending only on the marginal posterior model probabilities (i.e., there is a marginalizing effect). 

\begin{prop1}\label{prop1}
Suppose that RJMCMC proposals are of the form described in \eqref{eq:chaintrans}, and for each $k \in {\cal K}$, satisfy $T_{k}\sharp \, \pi_k = \otimes_{n_k} \nu$. Then,  \eqref{eq:rjacceptanceprobability} reduces to 
\begin{align}
    \alpha\big(\v x,\v x'\big) 
    &= 1 \wedge 
    \frac{\pi(k')}{\pi(k)}
    \frac{j_{k'}(k)}{j_{k}(k')}
    \label{eq:acceptancereduced2}.
\end{align}
\end{prop1}

\begin{corollary}\label{prop1corollary}
Provided the conditions of Proposition \ref{prop1} are satisfied, choosing $\{j_k\}$ such that
\begin{align}
     \pi(k')j_{k'}(k) =  \pi(k)j_{k}(k'), \quad \forall k,k' \in \mathcal{K},
\label{eq:optimalmodprop}
\end{align}
leads to a rejection-free proposal.
\end{corollary}

A natural choice when allowing {\em global} moves, i.e., $j_{k}(k') > 0$ for all $k,k' \in \mathcal{K}$ that solves the above detailed balance condition \eqref{eq:optimalmodprop}, is simply to choose
$j_k(k') = \pi(k')$ for all $k\in \mathcal{K}$. 
Of course, in practice the marginal probabilities $\pi(k)$ are typically unknown. However, the above observation is potentially instructive as to what a good choice may be if some approximation of model probabilities is available. 

\begin{tcolorbox}[colback=blue!10!white,colframe=blue!75!black]
Transdimensional RJMCMC proposals that employ {\em exact} transport maps to and from a reference distribution result in the same acceptance probabilities as performing {\em marginal} MCMC on the model indicator space and thus are in some sense an optimal proposal for general RJMCMC settings.  
\end{tcolorbox}

\section{NUMERICAL EXPERIMENTS}\label{sec:numerical}
For all experiments, we use masked autoregressive transforms with rational quadratic splines (RQMA) \cite[]{durkan_nflows_2020}. Prior to the RQMA transforms, a fixed (i.e., non-trainable) element-wise standardization is applied, which we find has a positive effect on the speed and stability of training. We also consider simpler NFs in the form of affine transformations, which are constructed using the sample mean vector and Cholesky factorization of the sample covariance matrix of the training set (Affine NFs). We highlight that in general, the choice of normalizing flow family is essentially arbitrary. However, as the proposed TRJ approach requires that the flows be evaluated in both directions (i.e., both the original flow and its inverse are needed in order to transform both to and from the reference distribution), it is computationally convenient to employ flows that are analytically tractable in both directions (as is the case for RQMA and Affine flows). Such tractability does not hold for all flows. For example, Invertible Residual Networks \citep{iresnet-behrmann19a} require numerical fixed-point iteration to evaluate in the inverse direction. In all experiments, flows are fit for each individual $k$-conditional target using a set of $N$ training samples obtained via a pilot run of either MCMC, Sequential Monte Carlo samplers \citep{del_moral_sequential_2006}, or exact sampling. The performance of transdimensional proposals is benchmarked in two ways, discussed below.

\paragraph{Model Probability (Running Estimates).} One way to benchmark the performance of an RJMCMC chain is to visualize the running estimates of the marginal posterior probability of a given model $k$ as the RJMCMC chain progresses. On average, a chain that mixes more rapidly is expected to produce more accurate results with fewer iterations.

\paragraph{Bartolucci Bridge Estimator.} To assess the effect of different transdimensional proposals alone (as opposed to in combination with within-model moves), we employ the Bayes factor bridge-sampling estimator of \citet[Eq.16]{bartolucci_efficient_2006}. The estimator relies on constructing a Rao-Blackwellized bridge sampling approach. An advantage of the approach as originally introduced is that the estimators can be computed from stored values obtained from a standard RJMCMC run. Specifically, the Bayes factor  
$B_{k,k'}$ (ratio of marginal likelihoods) is estimated via \begin{equation}
    \hat{B}_{k,k'} = \frac{N_{k'}^{-1}\sum_{i=1}^{N_{k'}} \alpha_{i}'}{N_k^{-1}\sum_{i=1}^{N_{k}}\alpha_{i}},
    \label{eq:Bartolucci}
\end{equation}
where $N_{k'}$ and $N_k$ are the number of proposed moves from model $k'$ to $k$, and from $k$ to $k'$, respectively in the run of the chain. The $\alpha$ and $\alpha'$ terms are simply the acceptance probabilities corresponding to those proposals, respectively. Rather than use the output of an RJMCMC run, a straightforward alternative when one has samples from the individual conditionals $\pi_k$ (not necessarily obtained via RJMCMC) is the following. First, for each sample from each target, make one proposal of RJMCMC (inclusive of the random choice of model to jump to). Then, treat each of these as if they were a proposal obtained from a standard RJMCMC chain, used together to compute estimators of the form \eqref{eq:Bartolucci}. In the special case when prior model probabilities are uniform, it is straightforward to convert estimators of Bayes factors to estimators of model probabilities \citep{bartolucci_efficient_2006} via
\begin{align}
    \hat{\pi}(k)&=\hat{B}_{j,k}^{-1}\bigg(1+\sum_{i\in\mathcal{K} \setminus \{j\}}\hat{B}_{i,j}\bigg)^{-1},
    \label{eq:BartoProbs}
\end{align}
for arbitrary $j\in\mathcal{K}$. In the sequel, the combination of multiple estimators of the form \eqref{eq:Bartolucci} within \eqref{eq:BartoProbs} is referred to as the {\em Bartolucci bridge estimator} (BBE) of model probabilities. We expect the BBE to exhibit the lowest variance for the best across-model proposal type when computed using the same input set of samples from all models (as well as the same model jump probabilities $\{j_k\}$).
When evaluating performance using the BBE, we generate an {\em additional} set of $N$ samples (for each model $k$) that are independent of the training samples. This set is referred to as the {\em evaluation} set. One proposal of RJMCMC is made using each of these evaluation samples to obtain the BBE \eqref{eq:Bartolucci} terms within the estimator of model probabilities \eqref{eq:BartoProbs}. 

\subsection{Illustrative Example}\label{sec:illustrativeex}
\begin{figure}[t]
\vspace{.2in}
\includegraphics[width=0.49\textwidth,trim={0 0 0 0.2cm},clip]{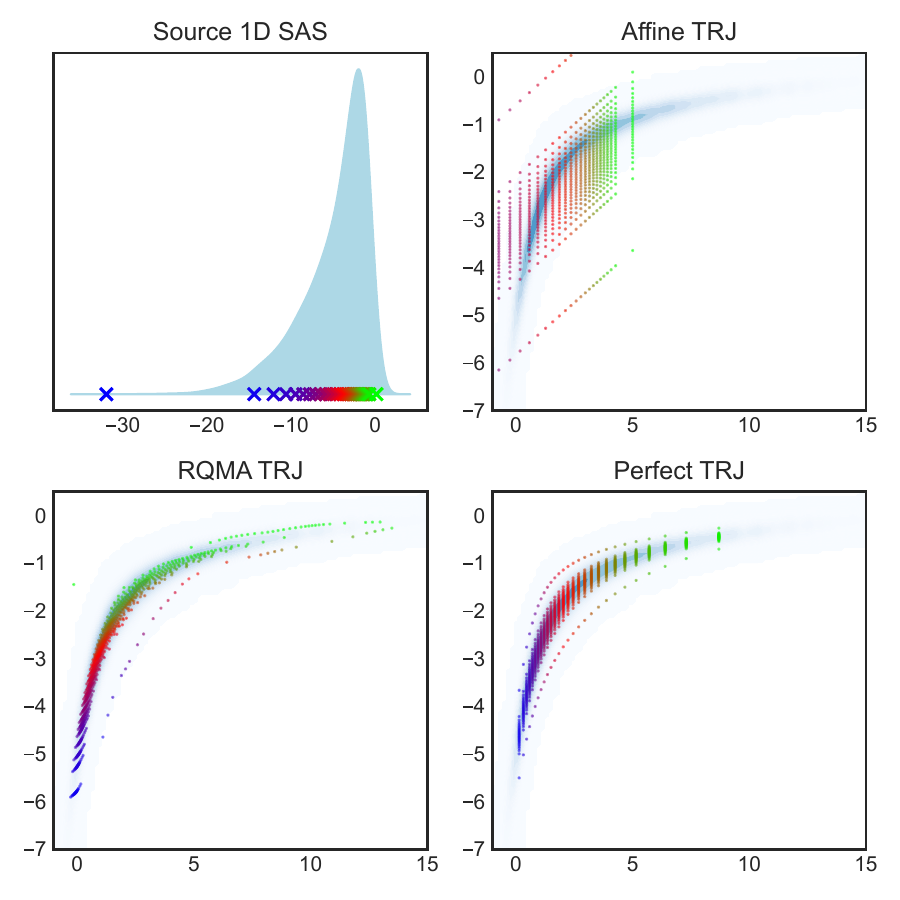}
\caption{Systematic draws from conditional target $\pi(x_1|k=1)$ defined in \eqref{eq:sastarget} are transported from $(1,\theta_1)\in\mathcal{K}\times\mathbb{R}^1$ (\emph{top left}) to $(2, (\theta_1,\theta_2))\in\mathcal{K}\times\mathbb{R}^2$ via proposals using: (1) approximate affine (\emph{top right}), (2) approximate RQMA-NF (\emph{bottom left}), and (3) perfect (\emph{bottom right}) TMs. The auxiliary variables in the proposals are also drawn systematically (30 for each point in the source distribution).  }\label{fig:sasproposals}
\end{figure}

\begin{figure}[h]
\centering
\includegraphics[width=0.45\textwidth]{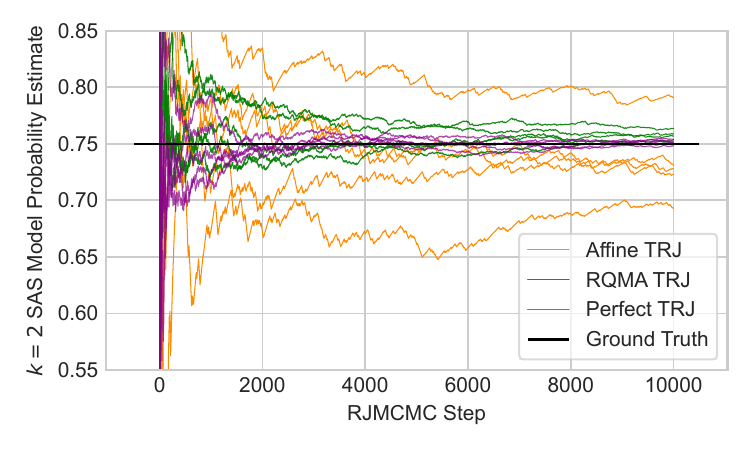}
\caption{Running estimates of the model probabilities for the $k=2$ component of the Sinh-Arcsinh target \eqref{eq:sastarget}. Proposals are all Algorithm \ref{algo:TRJ} with input approximate TMs (1) Affine, (2) RQMA-NF, (3) Perfect \eqref{eq:sas_perfectT}. Five chains on each proposal type are depicted, where alternating within-model proposals are a simple normal random walk.
}\label{fig:sas_trace}
\end{figure}

To construct an illustrative example where the exact TMs are known, we  use the (element-wise) inverse sinh-arcsinh (SAS) transformation \citep{jones_sinh-arcsinh_2009},
\[S_{\v \epsilon, \v \delta}(\cdot)= \sinh (\v\delta^{-1}\odot(\sinh^{-1}(\cdot)+\v \epsilon)),\]
where $\v \epsilon \in {\mathbb{R}}^n$, and $\v \delta \in \mathbb{R}_+^n$. For an $n \times n$ matrix $\m L$, define a transform $T(\v Z)$ where
\begin{align}
    T(\v Z) = S_{\v \epsilon, \v \delta}(\m L\v Z),
\end{align}
and $\v Z\sim\mathcal{N}(\v 0_n, \m I_{n\times n})$. The probability density function for the transformed variable $\v \theta = T(\v Z)$ is
\[
    p_{\v \epsilon, \v \delta, \m L}(\v \theta) =
    \phi_{\m L \m L^\top}\bigg( S_{\v \epsilon, \v \delta}^{-1}(\v \theta)\bigg) \big|J_{S_{\v \epsilon, \v \delta}^{-1}}(\v \theta)\big|,
\]
where $S^{-1}_{\v \epsilon, \v \delta}(\cdot)= \sinh(\v\delta \odot \sinh^{-1}(\cdot)- \v \epsilon)$. We consider a transdimensional target consisting of two models of dimension $n_1=1$ and $n_2=2$. 
We assign a probability of $1/4$ for model $k=1$ with parameters $\v \theta_1=(\theta_1^{(1)})$ and model probability of $3/4$ for $k=2$ with parameters $\v \theta_2 = (\theta_2^{(1)},\theta_2^{(2)})$. The target of interest for this example is
\begin{align}
    \pi(k,\v \theta_k)&=\begin{cases}
    \frac{1}{4}p_{\epsilon_1, \delta_1, 1}\big(\v \theta_{1} \big), & k=1, \\
    \frac{3}{4}p_{\v \epsilon_2, \v \delta_2, \m L}\big(\v \theta_2\big), & k=2,
    \end{cases}\label{eq:sastarget}
\end{align}
where 
\begin{equation}\begin{alignedat}{2}
    \epsilon_{1}&=-2,  
    &&\delta_{1}=1, \\
    \v\epsilon_{2}&=(1.5, -2), \ \ 
    && \text{ and } \ \v\delta_{2}=(1,  1.5), 
    \label{eq:sasparams}
\end{alignedat}\end{equation} 
and $\m L$ is a lower-triangular matrix such that 
\[ \m L \m L^\top = \begin{bmatrix} 1 & 0.99 \\ 0.99 & 1 \end{bmatrix}. \] 
By construction, for a reference distribution chosen to be a multidimensional standard Gaussian of appropriate dimension, an exact transport is given by the function 
\begin{align}
T^{-1}(\cdot) = \m L^{-1}  S_{\v \epsilon, \v \delta}^{-1}(\cdot).\label{eq:sas_perfectT}
\end{align}
Note that the choice of $j_{k}(k')= \frac{1}{4}\bb I_{\{k' = 1\}} + \frac{3}{4}\bb I_{\{k' = 2\}}$, for $k\in \{1,2\}$, i.e., the mixture weights,  correspond to a rejection-free proposal when the exact transports of the form in \eqref{eq:sas_perfectT}  are used. Both the affine and RQMA NF are trained on a set of $N=5\cdot 10^4$ samples obtained via exact simulation for each of the two targets. Figure \ref{fig:sasproposals} depicts a visualization of the proposal distributions (including those obtained using an exact transport map), and Figure \ref{fig:sas_trace} visualizes the running estimate of model probabilities for each proposal type, where $j_k(k')$ is set to the mixture weights as above. 
On average, the accuracy of the running model probability estimates is best for the exact TM, followed by RQMC NF, which provides a better approximation of the exact TM compared to Affine.


\subsection{Bayesian Factor Analysis Example}\label{sec:fa}

Here we consider the factor analysis (FA) model from \cite{lopes_bayesian_2004}. Factor analysis assumes the model $\v Y\sim\mathcal{N}(\v 0_6,\m \Sigma_{6\times 6})$ where the covariance matrix takes the form $\v\beta_k\v\beta_k^\top+ \Lambda$, where $\Lambda$ is a $6\times 6$ positive diagonal matrix, $\v\beta_k$ is a $6\times k$ lower-triangular matrix with a positive diagonal, $k$ is the number of factors, and the total number of parameters is $6(k+1)-k(k-1)/2$. We model monthly exchange rates of six currencies relative to the British pound, spanning January 1975 to December 1986 \cite[]{west_bayesian_1997} as 143 realizations, denoted as $\v y_i\in \mathbb{R}^6$ for $i=1,\ldots,143$. For this example, we are interested in $k=2$ or $3$ factors, yielding parameter spaces of dimensions 17 and 21 respectively. Following \cite{lopes_bayesian_2004}, for each $\v \beta_k=[\beta_{ij}]$ with $i=1,\ldots,6$, $j=1,\ldots,k$, the priors are
\begin{align}
\begin{split}
    \beta_{ij}&\sim\mathcal{N}(0,1),\ \ i < j,\\
    \beta_{ii}&\sim\mathcal{N}_{+}(0,1), \\ %half-normal
    \Lambda_{ii}&\sim\mathcal{IG}(1.1,0.05),\label{eq:faprior}
\end{split}
\end{align}
where $\mathcal{N}_+$ denotes the positive half-normal distribution and $\mathcal{IG}$ the inverse gamma distribution. We are interested in the posterior of $\v \theta_k = (\v \beta_k, \v \Lambda)$ for $k=2$ or $3$ factors. Via Bayes' Theorem, the posterior is
\begin{align}
    \pi(k,\v\theta_k| \v y) \varpropto p(k)p(\v\beta_k|k)p(\v\Lambda) \prod_{i=1}^{143} \phi_{\v \beta \v \beta^\top + \Lambda}(\v y_i),\label{eq:bayesposteriorfa}
\end{align}
where $\v y = (\v y_1,\ldots, \v y_{143})$. In this example, we compare TRJ proposals to the original RJMCMC proposal from \cite{lopes_bayesian_2004}. The form of the \cite{lopes_bayesian_2004} RJMCMC proposal is an independence proposal trained on the conditional posterior of each model. Write $\v \mu_{\v\beta_k}$ and $\m B_k$ as the posterior mean and covariance matrix of $\v\beta_k$, respectively. Defining $\v\theta_k=(\v\beta_k,\v\Lambda)$, the independence proposal is $q_k(\v\theta_k)=q_k(\v\beta_k)\prod_{i=1}^6 q_k(\Lambda_{ii})$, where for $k\in\mathcal{K}$, $q_k(\v\beta_k)=\mathcal{N}(\v \mu_{\v\beta_k},2\m B_k)$, and $q_k(\Lambda_{ii})=\mathcal{IG}(18,18\upsilon_{k,i}^2)$ where each $\upsilon_{k,i}^2$ is chosen to be an estimate of the marginal mode of $\Lambda_{ii}$ for model $k$ obtained from the training samples. 

To obtain training and evaluation sets of samples from \eqref{eq:bayesposteriorfa}, we use the No-U-Turn Sampler (NUTS) \cite[]{hoffman_no-u-turn_2014} as implemented in {\sf PyMC} \cite[]{salvatier_probabilistic_2016} for the three-factor model, and static temperature-annealed sequential Monte Carlo (SMC) \cite[]{del_moral_sequential_2006} for the two-factor model. The number of training and evaluation samples is kept equal, and the experiments consider both $N=2\cdot 10^3$ and $N=1.6 \cdot 10^4$ for each. Ten pairs of training and evaluation sample sets are generated. Each evaluation process produces a single BBE estimate, and this is conducted 100 times per training/evaluation pair, yielding a total of $10^3$ BBE estimates. 

\begin{figure}[h!]
\centering
\includegraphics[width=0.45\textwidth]{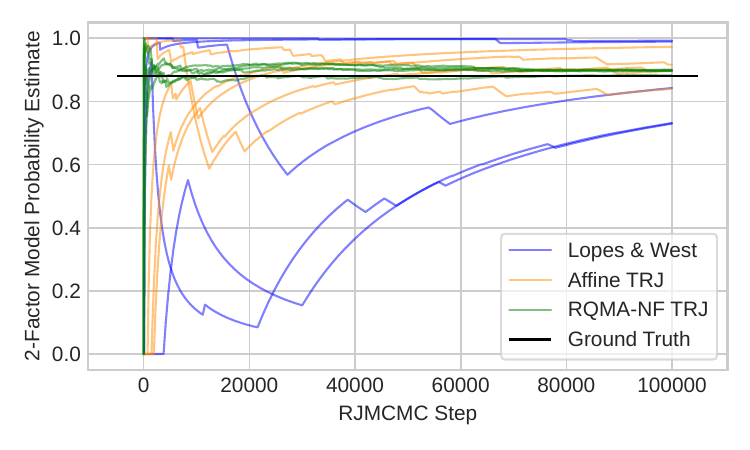}
\caption{Running estimates of the model probabilities for the two-factor model. Proposal types are the \cite{lopes_bayesian_2004} independence proposal, and Affine and RQMA-NF proposals using Algorithm \ref{algo:TRJ}. Five chains on each proposal type are depicted, where alternating within-model proposals are not shown.
}\label{fig:fa_trace}
\end{figure}

\begin{figure}[h!]
\centering
\includegraphics[width=0.475\textwidth]{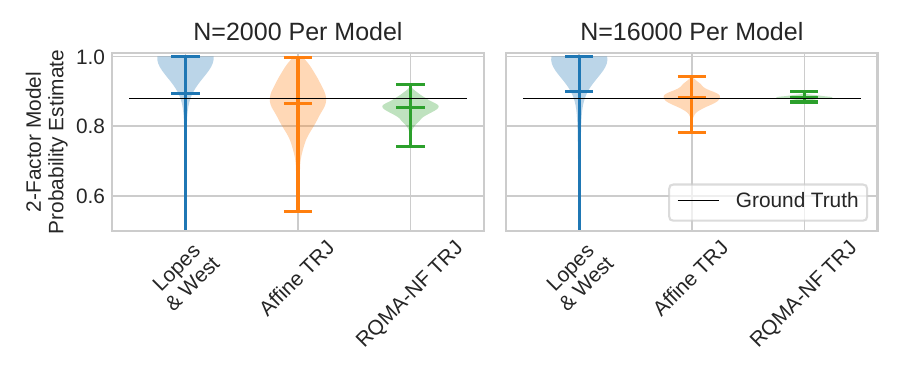}
\caption{Violin plot for the Factor Analysis example, showing the variability of the two-factor model probability estimates in the case where only the two-factor and three-factor models are compared. Model probability estimates are obtained via the BBE. 
}\label{fig:fa_bbe}
\end{figure}

\begin{figure}[!ht]
\centering
\includegraphics[width=0.45\textwidth]{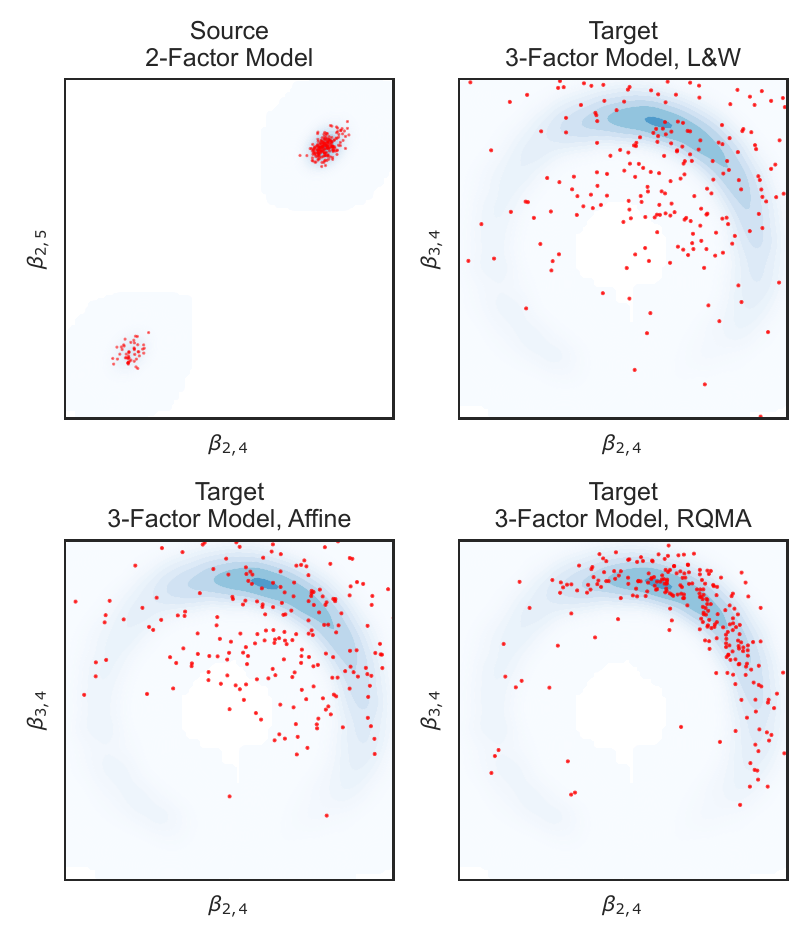}
\caption{A visualization (using selected bivariate plots) of the proposal from points on the 2-factor model (\emph{top-left}) to proposed points on the 3-factor model for each proposal type: \cite{lopes_bayesian_2004} independence RJMCMC proposal (\emph{top right}); TRJ with Affine map (\emph{bottom left}); TRJ with RQMA-NF map (\emph{bottom right}). Ground truth kernel density estimate was made using $5\cdot10^4$ SMC particles from \cite{south_2019_sequential}.} \label{fig:FAproposals}
\end{figure}

Figure \ref{fig:fa_trace} visualizes a comparison of the running estimate of the two-factor posterior model probability for RJMCMC chains that use the independence proposal and the RQMA-NF TRJ proposal against the ground truth value of $0.88$ as reported in \cite{lopes_bayesian_2004}. Each chain was run for $10^5$ steps with an RJMCMC proposal and a within-model random walk proposal used at each step in sequence. Figure \ref{fig:fa_bbe} visualizes the variability of model probability estimates obtained from the BBE. Both figures demonstrate the superior performance of the TRJ sampler approach with RQMA flows compared to the standard approach of \cite{lopes_bayesian_2004}. Figure \ref{fig:FAproposals} visualizes the transdimensional proposals with respect to selected bivariate marginals. 


\section{CONDITIONAL TRANSPORT PROPOSALS}\label{sec:cnfrj}
So far, we have introduced a proposition for rejection-free RJMCMC moves based on perfect TMs and model probabilities, and an approach to approximate both of the unknown quantities. However, there is one clear drawback to this approach: larger model spaces, such as those encountered in variable selection problems, would still require $k$ {\em individual} approximate TMs to be trained. This section explores how to overcome the above problem by modifying the sampling problem appropriately via a {\em saturated space} approach, so one can obtain the approximate TMs by training a single {\em conditional} NF (e.g., \cite{winkler_learning_2019}). The general idea of the
approach is that the model indicator is given as an input (context) to a function that outputs the parameters of a flow (or
sequence thereof). Such an approach implicitly defines a family of flows across {\em all} models, yet only requires training once. The benefit is thus primarily computational. We employ the notation $\tilde{T}(\cdot | k)$ for the flow obtained by providing the conditional normalizing flow model with context $k \in \mathcal{K}$.

The dimension-saturation approach, originally considered in \cite{brooks_efficient_2003}, is an equivalent formulation of RJMCMC that involves an augmented target. Writing the maximum model dimension as $n_{\rm max}$ (assumed to be finite) and recalling that $\v x=(\v\theta_k,k)$, we define our augmented target to be
\begin{align}
    \Tilde{\pi}(\Tilde{\v x})&=\pi(\v x)(\otimes_{n_{\rm max}-n_k}\nu)(\v u_{\backsim k}),\label{eq:refaugtarget}
\end{align}
where $\Tilde{\v x}=(k,\v\theta, \v u_{\backsim k})$, and ``$\backsim{k}$'' identifies that the auxiliary variable is of dimension $n_{\rm max}-n_k$. In this setting, one can obtain approximate TMs by training a single conditional NF with the conditioning vector being the model index $k\in \mathcal{K}$ (further details are provided in the supplement).
The associated proposals analogous to those in  \eqref{eq:chaintrans} are constructed as $(\v\theta_{k'}',\v u_{\backsim k'}) = c_{k'}^{-1}\circ \tilde{T}^{-1}(\cdot |k')\circ \tilde{T}(\cdot |k)\circ c_k(\v\theta_k,\v u_{\backsim k})$,
where $k'\sim j_{k}$, and $c_k$ is simply concatenation. The above yields Algorithm \ref{algo:ctrj}.

\begin{algorithm}[!ht]
\SetKwInOut{KwInput}{input}
\SetKwInOut{KwOutput}{output}
\small
\DontPrintSemicolon
\caption{CTRJ: Conditional Transport RJMCMC}\label{algo:ctrj}
\KwInput{$\Tilde{\v x} = (k,\v\theta_k,\v u_{\backsim k}),
 \Tilde{T},
\Tilde{\pi},
\{j_k\}, 
\nu, \{P_{k}(\v\theta, d\v\theta')\}$}
Draw $k' \sim j_{k}$\\
\uIf{$k'=k$}{

\text{Draw} $\v\theta_k'\sim P_{k}(\v\theta_k, d\v\theta_k')$\\
\KwRet{$(k,\v \theta_k',\v u_{\backsim k})$} 

}\uElse{
$\begin{aligned}
    \v\xi_k &\gets c_{k,k'}(\v\theta_k,\v u_{\backsim k})\\ 
    \v z&\gets \Tilde{T}(\v\xi_k|k)\\
    \v\xi_{k'} &\gets \Tilde{T}^{-1}(\v z|k')\\
    \v (\v\theta_{k'}',\v u_{\backsim k'})&\gets c_{k',k}^{-1}(\v\xi_{k'})\\
    \Tilde{\v x}'&\gets(k',\v\theta_{k'}',\v u_{\backsim k'})\\
    \alpha&\gets 1 \wedge \frac{\Tilde{\pi}(\Tilde{\v x}')}{\Tilde{\pi}(\Tilde{\v x})}\frac{j_{k'}(k)}{j_{k}(k')}
    \Big|J_{\Tilde{T}}(\v\theta_{k}|k)\Big|\Big|J_{\Tilde{T}}(\v\theta_{k'}'|k')\Big|^{-1}
\end{aligned}$\\
\vspace{.5em}
Draw $V\sim\mathcal{U}(0,1)$\\
\lIf{$\alpha>V$}{
\KwRet{$\Tilde{\v x}'$}
}\lElse{
\KwRet{$\Tilde{\v x}$}
}
}%
\end{algorithm}
\subsection{Variable Selection in Robust Regression Example}\label{sec:blockvs}
Here, a variable selection example is provided as a proof of concept for the conditional transport approach just described. For this example, we seek a challenging Bayesian model choice problem where conventional methods do not perform well. The interest is in realizations of a random response variable $\v Y = (Y_1,\ldots, Y_{80})^\top$, where for each $Y_i$, we have associated covariates $(x_{i0}, x_{i1}, x_{i2}, x_{i3})$ where $x_{i0}=1$. Writing $\m X$ for the associated $80\times 4$ design matrix, our model is $\v Y = \m X \v \beta + \epsilon$, 
where $\v \beta=(\beta_0,\ldots,\beta_3)$ is the vector of unknown regression parameters and $\v \epsilon = (\epsilon_1,\ldots,\epsilon_{80})$ is the residual error term. We model $\v \epsilon$ as a vector of independently and identically distributed random variables, each distributed as a mixture of a standard normal variable and a zero-mean normal variable with variance $10^2$.
To ease the exposition, we adjust the notation slightly to accommodate the Bayesian variable selection literature. Let $k$ be an indicator vector where the $i$th position of $k$ equals 1 if the variable $\beta_i$ is included in model $k$ and 0 otherwise --- e.g., the model indexed by $k =(1,0,1,1)$ corresponds to the model with parameters $\v \theta_k = (\beta_0, \beta_2, \beta_3)$. We consider a model space of cardinality of 4, composed of all models of the form $k = (1,k_1,k_2,k_2)$ where $k_i\in \{0,1\}$ for $i=1,2$. These models will include or exclude the parameters $\beta_2$ and $\beta_3$ jointly as commonly considered in group variable selection (e.g. \cite{xu_bayesian_2015}). The prior distributions are specified as
\begin{align}
\begin{split}
k_i&\sim\mathrm{Bernoulli}(\sfrac{1}{2} ),\quad i\in\{1,2\}, \quad \text{ and }\\ 
\beta_i&\sim\mathcal{N}(0,10^2),\quad i\in\{0,1,2,3\}.
\end{split}
\end{align}
The target distribution $\pi$ is then the posterior distribution over the set of models and regression coefficients, obtained via Bayes' Theorem.

The dataset for the example was simulated from the model $k=(1,1,0,0)$, which has  corresponding parameters $\v \theta_k = (\beta_0,\beta_1)$. 
We simulated half of the observations of the response using $(\beta_0, \beta_1) = (1,1)$ and half using $(\beta_0,\beta_1)=(6,1)$. This choice was made to induce further challenging multi-modal features in the resulting posterior. The elements of the matrix $\m X$ were simulated independently from a standard normal distribution, and the elements of the residual vector $\epsilon$ were simulated from a normal distribution with a mean of zero and a variance of 25. See the supplementary material for visualizations of the multimodal posterior behavior. The design of each proposal is as follows. The ``standard'' proposal adopts the {\em independent auxiliary variable} method of \cite{brooks_efficient_2003}, which is equivalent to using the augmented target \eqref{eq:refaugtarget} where $\nu$ is the prior distribution for $\beta_i$, i.e., $\mathcal{N}(0,10^2)$. The standard proposals are deterministic in the sense that no auxiliary variables are drawn, but rather the $i^{\rm th}$ block of parameters is reinstated to its previous value when $k_i=1$. The TRJ affine and RQMA-NF proposals apply Algorithm \ref{algo:TRJ} with approximate affine, and RQMA-NF TMs respectively trained on samples from the conditional target distributions. The CTRJ proposal adopts the augmented target \eqref{eq:refaugtarget} where $\nu$ is set to the reference univariate density as described in Section \ref{sec:cnfrj}, and proceeds as per Algorithm \ref{algo:ctrj}. The empirical performance of each proposal is examined for training and evaluation sets via the BBE. The number of training and evaluation samples was kept the same, and we considered both $N=500$ and $N=4000$ sampled from a single SMC run. The process of training the flow and estimation of model probability via BBE was repeated eighty times.

\begin{figure}[t]
\centering
\includegraphics[width=0.48\textwidth]{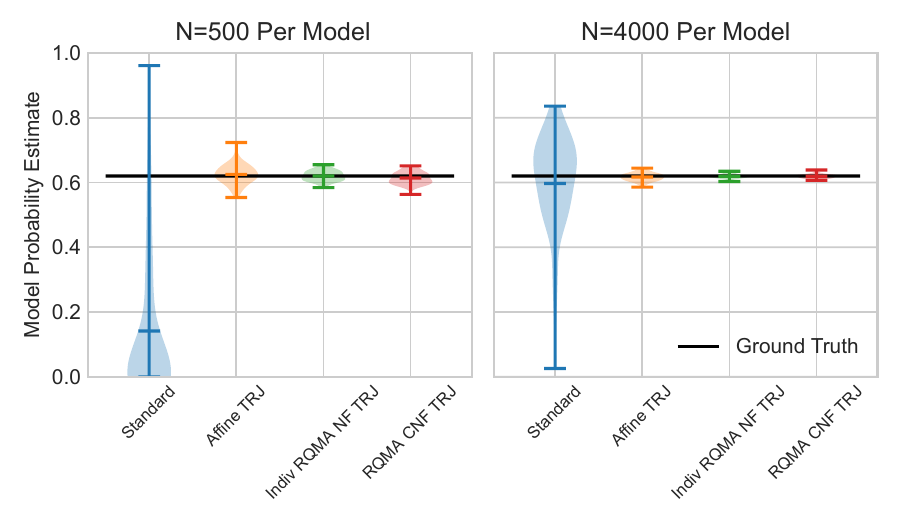}
\caption{Violin plot (for the example in Section \ref{sec:blockvs}) showing the variability of the $k=(1,1,1,1)$ model probability estimate for each proposal type using the Bartolucci Estimator vs a ground truth given by individual SMC ($N=5\cdot 10^4$). Here, Indiv RQMA-NF TRJ corresponds to Algorithm 1 with four {\em individual} normalizing flows, whereas RQMA-CNF TRJ corresponds to Algorithm 2 using a single {\em conditional} normalizing flow.}\label{fig:vs_bbe}
\end{figure}

Figure \ref{fig:vs_bbe} visualizes the variability of the $k=(1,1,1,1)$ model probability estimate using the same approach as in Section \ref{sec:fa}. The figure shows the vastly superior accuracy and variability of model probability estimates using the TRJ proposals relative to the independent auxiliary method. The performance of the conditional RQMA flow appears similar to that of the individual RQMA and affine flows but only requires fitting a single conditional flow as opposed to a flow per model. 

\begin{tcolorbox}[colback=blue!10!white,colframe=blue!75!black]
Numerical examples demonstrate a clear benefit to using transport RJMCMC proposals in cases where the transport maps are {\em approximate}, which is typically the best one can hope for in practice.
\end{tcolorbox}
\section{DISCUSSION}\label{sec:discussion}
The work introduces a method to design across-model RJMCMC proposals using ideas from measure transport. The approach was shown to have desirable properties in the case where exact TMs are used, while still providing good performance in the case where the TMs are approximate. Although some form of initial effort to obtain samples from each conditional target is somewhat common when calibrating proposals  (e.g., as in \cite{lopes_bayesian_2004, hastie_towards_2005}), this is still a weakness of TRJ proposals, along with the  computational effort required to train the approximate transport maps. The use of conditional NFs may mitigate the latter somewhat, but a promising avenue for future research would be some form of amortized approach in model spaces where there is some shared structure (and thus perhaps samples from only {\em some} models would suffice).   Moreover, the effort may be justified in settings where likelihoods are particularly expensive, gradient-based methods are unavailable (e.g., implicit models), or by and large the gain in performance appropriately justifies the effort in obtaining the approximate TMs. Finally, while we mainly repurposed the BBE to benchmark cross-model proposal quality, the results seem promising, and it may be interesting to explore its use in lieu of standard RJMCMC if efforts to obtain approximate TMs have already occurred. 
\subsubsection*{Acknowledgements}
Christopher Drovandi acknowledges support from an Australian Research Council Discovery Project (DP200102101), which Laurence Davies was also partially supported by. 

\subsubsection*{References}
\vspace{-2em}
\bibliographystyle{apalike}
\bibliography{references_filtered}
\appendix

%% file: supplement_body.tex
\onecolumn
\allowdisplaybreaks
\aistatstitle{SUPPLEMENTARY MATERIAL}
\appendix
%\nopagebreak[4]
% \section{Proof of Volume Preservation}
% Let $d=\max\{n_k,n_{k'}\}$ and $\v z\in\mathcal{Z}\subseteq\mathbb{R}^{d}$. Given a diffeomorphism $\bar{h}_{k,k'}:\mathcal{Z}\rightarrow\mathcal{Z}$ where the pushforward measure is $\bar{h}_{k, k'}\sharp \nu_{d} = \nu_{d}$, we note the inverse pushforward is $\bar{h}_{k, k'}^{-1}\sharp \nu_{d} = \nu_{d}$. Noting the identity $\bar{h}_{k,k'}(\mathcal{Z})=\bar{h}_{k,k'}^{-1}(\mathcal{Z})=\mathcal{Z}$, by the change of variables theorem given $\varphi$ is a measurable function on $\mathcal{Z}$ we have
% \begin{align}
%     \int_{\mathcal{Z}} \varphi d\lambda_d 
%     &=
%     \int_{\mathcal{Z}} (\varphi\circ\bar{h}_{k,k'})\big|J_{\bar{h}_{k,k'}}\big|d\lambda_d,
% \end{align}
% and
% \begin{align}
%     \int_{\mathcal{Z}} \varphi d\lambda_d 
%     &=
%     \int_{\mathcal{Z}} (\varphi\circ\bar{h}_{k,k'}^{-1})\big|J_{\bar{h}_{k,k'}^{-1}}\big|d\lambda_d.
% \end{align}
% The integrands on the right sides are equal, i.e.
% \begin{align}
%     (\varphi\circ\bar{h}_{k,k'})\big|J_{\bar{h}_{k,k'}}\big|
%     &=
%     (\varphi\circ\bar{h}_{k,k'}^{-1})\big|J_{\bar{h}_{k,k'}^{-1}}\big|
% \end{align}
\section{PROOF OF PROPOSITION 1}
\label{sec:prop1proof}
\begin{proof}
Here, we employ the shorthand notation $\nu_n := \otimes_n \nu$ and  $\nu_n(\cdot) := (\otimes_n \nu)(\cdot)$ for measure and probability density functions, respectively. For a diffeomorphism $(\v\theta_{k'}',\v u_{k'})=h_{k,k'}(\v\theta_k,\v u_k)$, auxiliary variables $\v u_k\sim g_{k,k'}$, $\v u_{k'}\sim g_{k',k}$ of respective dimensions $w_k$ and $w_{k'}$, if the dimension matching requirement $n_{k'} + w_{k'} = n_k + w_k$ is satisfied, then the standard choice \citep{green_reversible_1995} for the acceptance probability that satisfies detailed balance is
\begin{align}
    \alpha(\v x,\v x')&=1\wedge\frac{\pi(\v x')j_{k'}(k)g_{k',k}(\v u_{k'})}{\pi(\v x)j_{k}(k')g_{k,k'}(\v u_k)}\big|J_{h_{k,k'}}(\v\theta_k,\v u_k)\big|.\label{eq:proof_rjacceptanceprobability}
\end{align}
First, consider the case that $n_{k'} > n_k$. 
The proposed move from index $k$ to $k'$ is
\begin{equation}
\v \theta_{k'}' = h_{k,k'}(\v \theta_k, \v u_k) =  T_{k'}^{-1}(\bar{h}_{k,k'}(T_k(\v \theta_k), \v u_k)), \quad \text{ where } \v u_k \sim \nu_{w_k}.  \label{eq:proposed_move_identity} 
\end{equation}
In this setting, we have $\v u_{k'}=\emptyset$, $w_{k'}=0$, and $g_{k,k'}=\nu_{w_k}$. Note that $g_{k,k'} = \nu_0$ (i.e., an ``empty'' distribution), so $g_{k',k}(\v u_{k'})=1$. Further, the dimension-matching requirement reduces to $n_{k'} = n_k + w_k$.  Using the properties that $|J_{f_1 \circ f_2}| = |J_{f_1}|\cdot |J_{f_2}|$ and $|J_{f^{-1}}|=|J_{f}|^{-1}$, along with the assumption that $\bar{h}_{k,k'}$ is a volume-preserving map, i.e., $ \big|J_{\bar{h}_{k,k'}}\big|=1$, we have that 
\begin{align}
%\begin{split}
    \big|J_{h_{k,k'}}(\v\theta_{k'}',\v u_{k'})\big|=\big|J_{T_{k}}(\v\theta_{k})\big|\cdot\big| J_{\bar{h}_{k,k'}}(T_k(\v\theta_{k'}'),\v u_{k'})\big|\cdot\big|J_{T_{k'}}(\v\theta_{k'}')\big|^{-1}
    = \frac{\big|J_{T_{k}}(\v\theta_{k})\big|}{\big|J_{T_{k'}}(\v\theta_{k'}')\big|}.\label{eq:jac_h}
\end{align}
Making the above substitutions into \eqref{eq:proof_rjacceptanceprobability}, and employing the conditional factorization identity $\pi(\v x) = \pi(k)\pi_k(\v \theta_k)$ to both $\pi(\v x)$ and $\pi(\v x')$ yields
\begin{align}
    \alpha(\v x,\v x')&=1\wedge\frac{j_{k'}(k)\pi(k') }{j_{k}(k')\pi(k)}\left(\frac{\pi_{k'}(\v \theta_{k'}')\big|J_{T_{k}}(\v\theta_{k})\big|}{\nu_{w_k}(\v u_k)\pi_k(\v \theta_k)\big|J_{T_{k'}}(\v\theta_{k'}')|}\right).\label{eq:proof_rjacceptanceprobability2}
\end{align}
Hence, showing that the bracketed term above must always be equal to one establishes the desired result. To accomplish this, we make use of several identities. 
First, note that by hypothesis, $T_k$ and $T_{k'}$ are diffeomorphic and satisfy $T_{k}\sharp\pi_k = \nu_{n_k}$ and $T_{k'}\sharp\pi_{k'} = \nu_{n_{k'}}$. Thus, by the standard change of variables theorem, we obtain the identities
\begin{align}
\begin{split}
    \pi_k(\v\theta_k) = \nu_{n_k}\big(T_{k}(\v\theta_k)\big)\big|J_{T_{k}}(\v\theta_{k})\big|, \quad \text{ and } \quad   \pi_{k'}(\v\theta_{k'}') = \nu_{n_{k'}}\big(T_{k'}(\v\theta_{k'}')\big)\big|J_{T_{k'}}(\v\theta_{k'}')\big|. \label{eq:condtarget_identity}
\end{split}
\end{align}
Next, by hypothesis $\bar{h}_{k,k'}\sharp \nu_{n_{k}+w_k} = \nu_{n_k + w_k}$, and since $\bar{h}_{k,k'}$ is a volume-preserving diffeomorphism we obtain
\begin{equation} \nu_{n_k'}(\v z) = \nu_{n_k'}(\bar{h}_{k,k'}^{-1}(\v z)).\label{eq:reference_h_identity}
\end{equation}
Finally, observe
\begin{equation}\begin{split} \bar{h}_{k,k'}^{-1} \circ T_{k}(\v \theta_{k'}') & \underbrace{=}_{\textstyle \eqref{eq:proposed_move_identity}}  \bar{h}_{k,k'}^{-1} \circ T_{k'} \circ  T_{k'}^{-1} \circ \bar{h}_{k,k'} (T_k(\v \theta_k), \v u_k) =  (T_k(\v \theta_k), \v u_k),
\end{split}\label{eq:longidentity}
\end{equation} 
where the final equality above uses  associativity of function composition. Using the above identities, and that $\nu_{n_{k'}}=\nu_{n_k + w_k}$, we obtain   
\[\frac{\pi_{k'}(\v \theta_{k'}')\big|J_{T_{k}}(\v\theta_{k})\big|}{\nu_{w_k}(\v u_k)\pi_k(\v \theta_k)\big|J_{T_{k'}}(\v\theta_{k'})\big|} \underbrace{=}_{\textstyle \eqref{eq:condtarget_identity}} \frac{\nu_{n_k'}(T_{k'}(\v \theta_k'))}{\nu_{n_k + w_k}(T_k(\v \theta_k), \v u_k)} \underbrace{=}_{\textstyle \eqref{eq:reference_h_identity}} \frac{\nu_{n_k'}(\bar{h}_{k,k'}^{-1}(T_{k'}(\v \theta_{k'}'))}{\nu_{n_k + w_k}(T_k(\v \theta_k), \v u_k)}\underbrace{=}_{\textstyle \eqref{eq:longidentity}}\frac{\nu_{n_k + w_k}(T_{k}(\v \theta_k), \v u_k)}{\nu_{n_k + w_k}(T_k(\v \theta_k), \v u_k)} = 1, 
\]
and the result is established for the stated case. The result is established for the case $n_k = n_{k'}$ by repeating the above argument but setting $\nu_{w_k}(\cdot) = 1$ because $w_k = 0$ in that case. Similarly, for the case where $n_{k'} < n_k$, repeat the above argument by applying the reverse move (replacing $h_{k,k'}$ with $h_{k',k} := h^{-1}_{k,k'}$), and using $\bar{h}^{-1}_{k',k}\sharp \nu_{n_k + w_k} = \nu_{n_k + w_k}$.
\end{proof}

\section{NORMALIZING FLOW CONFIGURATIONS}
This section gives details regarding the flows used in the experiments. Write $h:\mathbb{R}^n\rightarrow[0,1]^n$ to denote the  sigmoid function (applied element-wise). Define 
 \begin{equation}
 s(\v z; \v a, \v b) = \v b \odot (\v z - \v a), \quad \v a \in \bb R^n, \v b \in \bb R^n_+.   \label{eq:sup_flowconfig}  
 \end{equation}
The flow-based models used are of the form $T =  h^{-1}\circ F \circ h \circ s$, 
 where $F:[0,1]^n\rightarrow [0,1]^n$ is the composition of three (3) {\em masked autoregressive rational quadratic spline} transforms, implemented in the Python package {\sf nflows}. 
Parameters used for each of the three transforms were $10$ bins (corresponding to $11$ ``knot'' points), the autoregressive neural networks used in each transform had two hidden layers, each with $32 \times n$ hidden features. A Stochastic Gradient Descent (SGD) algorithm is used to train the parameters of the sequence of transforms $F$. The transform $s$ is fixed at the beginning of training and remains so during training, with the values of $\v a$ and $\v b$ taken to be the marginal means and reciprocal standard deviations of the data (samples), respectively, from a given $\pi_k$. It is beneficial to fix $s$ as described above as it essentially standardizes the data prior to model training. 

\subsection{Conditional Normalizing Flow}
The conditional approach uses the conditional transform
$\tilde{T}(\, \cdot \, | k) =  h^{-1}\circ F(\, \cdot \, | k) \circ h \circ s(\, \cdot \,  | k)$. In the previous equation, $F(\cdot ; k)$ is now a conditional normalizing flow model that, given as input a context (conditioning) variable $k$, outputs a flow of the same form as $F$ in the non-conditional case. Additionally, the CNF base distribution is now a {\em conditional} base --- an independent Gaussian whose mean and scale parameters are determined via a context $k$ (implemented as  {\em ConditionalDiagonalNormal} in {\sf nflows}). Finally, $s(\cdot | k)$ is a {\em conditional standardization} that takes into account the known distribution of the auxiliary variables as well as when they feature for different models. Note that under the augmented target distribution, conditional on $k \in {\cal K}$, $\v\theta_k$ is distributed according to the conditional target $\pi_k$, and the auxiliary variables $\v u_{\backsim k}$ are distributed according to $\nu_{n_{\rm max}-n_k}$.
Rather than simply standardize the auxilliary variables (which are known to be marginally distributed according to $\nu$) for each model, it is often straightforward to apply a perfect {\em Gaussianizing} transform.
Write $F_{\nu}$ for the cumulative density function (cdf) of the reference distribution $\nu$, and $\Phi^{-1}$ for the inverse cdf of the univariate standard normal distribution (i.e., the probit function). The generalization of the standardizing transform to the saturated state space (defined for output $i$) is then  
\begin{align}
    s(\v v|k ; \{\v a_k, \v b_k \}_{k \in \cal K})_i
    &=
    \begin{cases}
    b_{ki}\big(v_i - a_{ki}\big), & v_i \text{ is {\em not} an auxilliary variable for model } k,\\
    \Phi^{-1} \circ  F_{\nu}(v_i), & \text{otherwise}.
    \end{cases}
\end{align}
Similar to the unconditional case, the vectors $\v b_{k}$ and $\v a_k$ are fixed to be the reciprocal standard deviation and mean vectors of the data (samples) from each $\pi_k$. Optionally, one can set $\v a_k = \v a$ and $\v b_k = \v b$ for all $k$ using the same quantities obtained from pooling the samples (this is the approach implemented). In the considered example, the reference distribution $\nu$ was taken to be a standard Gaussian, and thus the second case above reduces simply to returning $v_i$. We highlight that the check for which case to apply in the function above is implemented via a {\em masking} function that, for input $k \in {\cal K},i\in\{1,\ldots, n_{\max} \}$, returns one if the condition is satisfied, and zero otherwise. 

\section{BLOCK VARIABLE SELECTION EXAMPLE}
Figure \ref{fig:vsposterior} depicts the distributions for each conditional target in the example. Note that each is multi-modal.  However, there are overlaps in some regions of  high probability mass, which could possibly assist the performance of the CTRJ proposal. Figure \ref{fig:vs_bbe_all} is an extended version of Figure \ref{fig:vs_bbe} in the paper, showing results of the Bartolucci estimator across varying number of samples $N$ for each of the four estimated model probabilities.

\setcounter{figure}{7}
\begin{figure}[h]
\centering
\includegraphics[width=.8\textwidth]{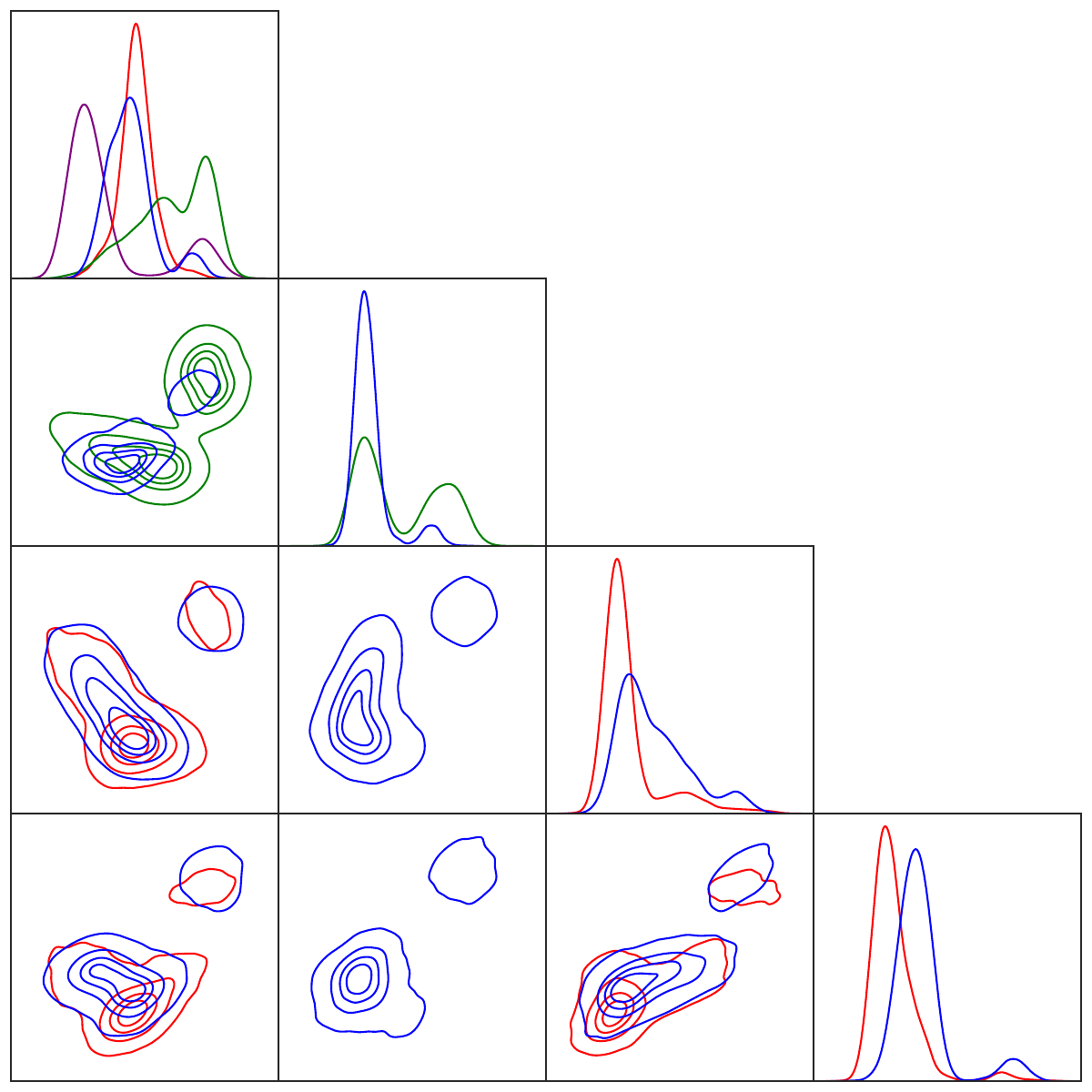}%
\caption{Pairwise plot of the conditional bivariate posterior densities in the Bayesian variable selection example. All four models feature: $k=(1,0,0,0)$ (\emph{Purple}) , $k=(1,1,0,0)$ (\emph{Green}), $k=(1,0,1,1)$ (\emph{Red}), and $k=(1,1,1,1)$ (\emph{Blue}).}
        \label{fig:vsposterior}
\end{figure}
\begin{figure}[h!]
\centering
\includegraphics[width=.9\textwidth]{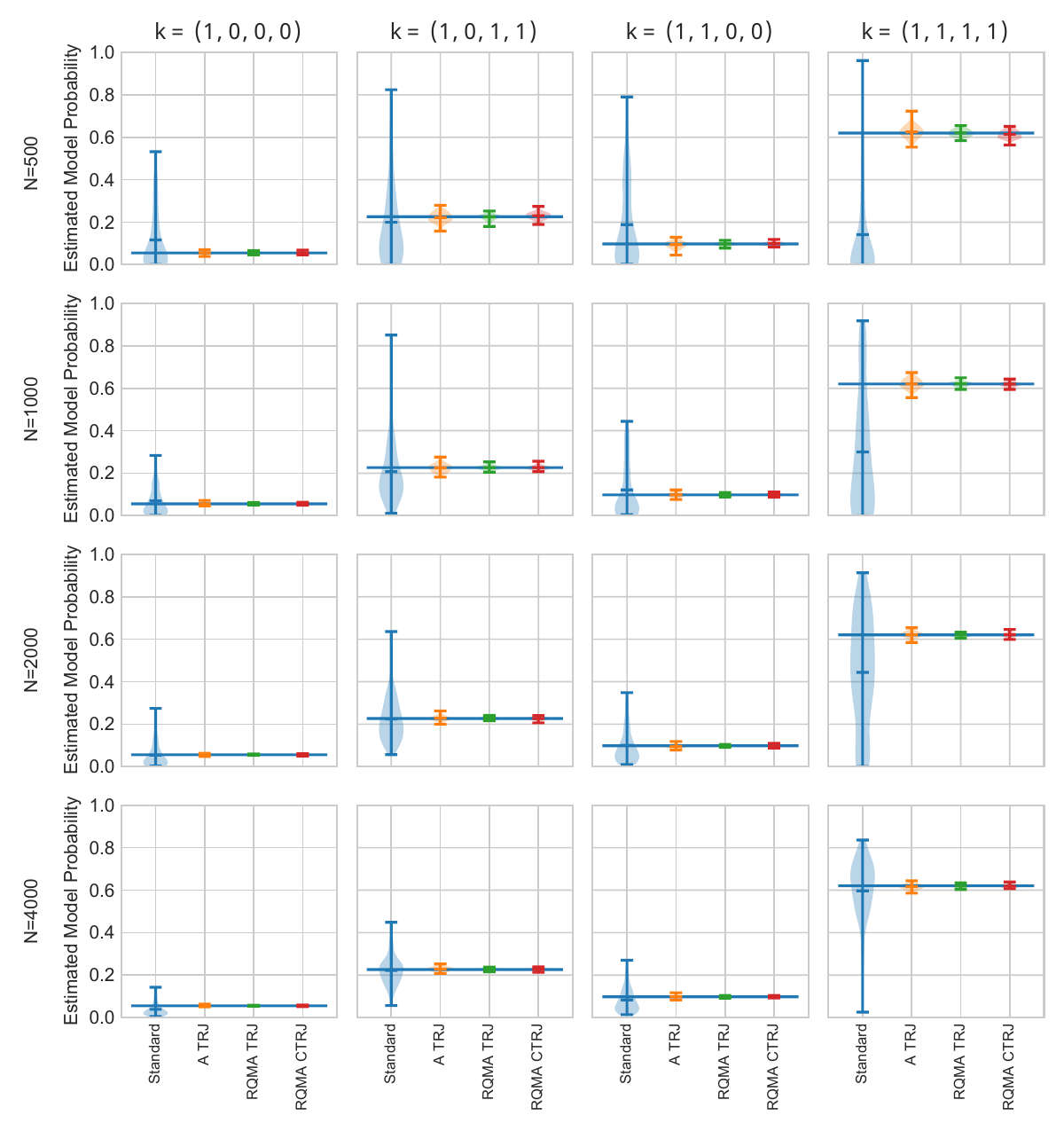}
\caption{Violin plot showing variability of each proposal type using the Bartolucci Bridge Estimator for the Bayesian variable selection example. To obtain samples to train the approximate transport maps, each model posterior was sampled using individual SMC samplers with $N \in \{500, 1000, 2000, 8000\}$ particles with a random walk kernel. Horizontal lines indicate ground-truth values obtained using large-sample runs of individual SMC ($N=5\cdot 10^4$).}\label{fig:vs_bbe_all}
\end{figure}